\documentclass{PoS}

\usepackage{amssymb}  
\usepackage{amsfonts}
\usepackage{amsmath}

\newcommand{\be}{\begin{equation}}
\newcommand{\ee}{\end{equation}}
\newcommand{\bea}{\begin{eqnarray}}
\newcommand{\eea}{\end{eqnarray}}
\newcommand{\ba}[1]{\begin{array}{#1}}
\newcommand{\ea}{\end{array}}
\newcommand{\non}{\nonumber\\}

\newcommand{\Tr}{{\rm Tr}}

\title{Relativistic superfluid hydrodynamics from field theory}

\ShortTitle{Relativistic superfluid hydrodynamics from field theory}

\author{Mark G.\ Alford\\
Department of Physics, Washington University St Louis, MO, 63130, USA\\
       E-mail: \email{alford@wuphys.wustl.edu}}

\author{S.\ Kumar Mallavarapu\\
Department of Physics, Washington University St Louis, MO, 63130, USA\\
        E-mail: \email{kumar.s@go.wustl.edu}}

\author{\speaker{Andreas Schmitt}\\
Institut f\"{u}r Theoretische Physik, Technische Universit\"{a}t Wien, 1040 Vienna, Austria\\
        E-mail: \email{aschmitt@hep.itp.tuwien.ac.at}}

\author{Stephan Stetina\\
Institut f\"{u}r Theoretische Physik, Technische Universit\"{a}t Wien, 1040 Vienna, Austria\\
        E-mail: \email{stetina@hep.itp.tuwien.ac.at}}

\abstract{It is well known that the hydrodynamics of a zero-temperature superfluid can be formulated in field-theoretic terms, relating for example
the superfluid four-velocity to the gradient of the phase of a Bose-condensed scalar field. At nonzero temperatures, where the phenomenology of a 
superfluid is usually described within a two-fluid picture, this relationship is less obvious. For the case of a uniform, dissipationless superfluid
at small temperatures and weak coupling we discuss this relationship within a $\varphi^4$ model. For instance,
we compute the entrainment coefficient, which describes the interaction between the superfluid and the normal-fluid components, and the velocities
of first and second sound in the presence of a superflow. 
Our study is very general, but can also be seen as a step towards understanding the superfluid properties of various phases of 
dense nuclear and quark matter in the interior of compact stars.}

\FullConference{Xth Quark Confinement and the Hadron Spectrum,\\
		October 8-12, 2012\\
		TUM Campus Garching, Munich, Germany}

\begin{document}

\section{Introduction and summary}

On a microscopic level, superfluidity arises through the formation of a Bose condensate which carries charge associated to a $U(1)$ symmetry of the 
underlying Lagrangian. The condensed bosons may be the fundamental degrees of freedom of the microscopic theory or may be composite objects such 
as Cooper pairs of fermions. In either case, the existence of the condensate allows for frictionless transport of the respective charge (mass, in the nonrelativistic
context). The reason is that, for zero temperature and for sufficiently small velocities of this ``superflow'', there are no elementary excitations
that could dissipate energy. This may seem surprising because the Bose condensate spontaneously breaks a global symmetry, giving rise to 
a massless mode -- the Goldstone mode -- that can potentially be excited. However, because the low-energy dispersion relation of the Goldstone mode is typically linear 
in momentum (and not quadratic or higher order), it turns out that it can only be excited beyond a certain critical velocity which equals the slope of the 
linear part of the dispersion. 

At nonzero temperature, on the other hand, excitations of the Goldstone mode (the ``phonons'') are present for any superfluid velocity, 
and the picture becomes more complicated. Our microscopic description of this situation will be as follows. We shall consider a complex scalar field $\varphi$ with 
a quartic interaction term $\lambda\varphi^4$. 
Then we consider the thermodynamics of this system in the presence of three independent, external parameters: chemical potential $\mu$, 
superfluid velocity ${\bf v_s}$, and temperature $T$. (We will see that $\mu$ and ${\bf v}_s$ are related to the temporal and spatial derivatives of the 
phase of the condensate, respectively. In the two-fluid picture, to be introduced below, our microscopic calculation is performed in the rest frame of the normal fluid, 
i.e., ${\bf v_s}$ is the velocity of the superfluid measured in the normal-fluid rest frame.)  We restrict ourselves to a uniform velocity ${\bf v}_s$ and to small
temperatures $T\ll \mu$ and small couplings $\lambda\ll 1$. 
This allows us to evaluate the effective action, the stress-energy tensor etc.\  analytically. Our approach naturally contains the effect of both the condensate and 
the phonons -- and we shall see how both are coupled to each other. It also contains in principle the full spectrum of the Goldstone mode, i.e., beyond the linear term mentioned above.
[For the explicit results we shall truncate the temperature expansion at $T^6$, for which only the linear and cubic parts of the dispersion are needed, see Eqs.\ (\ref{c1c2}).] 
Due to the nonzero superfluid velocity we obtain an anisotropic dispersion relation for the phonons. Although technically a bit involved, this approach is 
completely straightforward from the field-theoretic point of view. The main goal is to relate it to the usual ``macroscopic''
description of a superfluid, that is less common for field theorists but widely used in the phenomenology and hydrodynamics of superfluidity. We briefly 
explain this description now.

On a macroscopic level, a superfluid at nonzero temperatures (below the critical temperature where superfluidity breaks down) is usually described 
in the two-fluid model \cite{1938Natur.141..913T}, which was developed in the non-relativistic context of superfluid helium. In the two-fluid model, 
the superfluid is viewed as consisting of two fluid components that interpenetrate and entrain each other. Non-relativistically,
each of the fluid components is characterized by a mass density such that the sum of both mass densities adds up to the total (conserved) mass density
$\rho = \rho_s+\rho_n$.
While the superfluid density $\rho_s$ equals the total density at zero temperature, it decreases with temperature until all mass sits in the normal density $\rho_n$ 
at and above
the critical temperature. One of the predictions of this formalism is a second sound mode in superfluids because, besides the usual (first) sound -- 
oscillations in the total density (pressure) -- relative oscillations between the two fluid components become possible which correspond to oscillations in temperature 
(entropy). We are interested in the relativistic version of this two-fluid formalism \cite{1982PhLA...91...70K}. Here, the basic variables 
are four 4-vectors, namely the conserved current $j^\mu$, the entropy current $s^\mu$ (also conserved in the absence of dissipation), and their conjugate 
momenta, $\partial^\mu\psi$ and $\Theta^\mu$
(the connection to the microscopic description becomes manifest through the phase of the condensate $\psi$). 
Two of these 4-vectors are independent, i.e., we may for instance 
choose the two currents as our variables and obtain the momenta as linear combinations of them,
\bea\label{psiTheta}
\partial^\mu\psi &=&   B j^\mu +  A s^\mu  \, , \qquad \Theta^\mu =  A j^\mu + C s^\mu  \, .
\eea
The coefficients $A$, $B$, $C$ depend on the details of the system, i.e., the underlying microscopic theory. They can be viewed as derivatives of a 
generalized energy density $\Lambda$ with respect to Lorentz scalars built from the currents, see Eq.\ (\ref{ABC}). We shall show how to compute these 
coefficients within our scalar field theory. 
In general, the momenta will not be 4-parallel to their corresponding currents, but receive contributions also from the other current. This is interpreted as 
entrainment between the two fluid components and the corresponding coefficient $A$ is called entrainment coefficient. Note that here the two fluid components are given by 
the conserved current and the entropy current, not by the superfluid and the normal fluid. An equivalent formulation \cite{Son:2000ht}, somewhat closer to the 
non-relativistic version mentioned above, uses as independent variables one of the currents, namely $s^\mu$, which corresponds to the normal fluid, and one of 
the momenta, namely $\partial^\mu\psi$, which is proportional to the superfluid velocity \cite{Son:2000ht}. Both formulations can be translated into each 
other \cite{Alford:2012vn}.

At first sight, the hydrodynamical variables of the two-fluid formalism seem quite different from the variables of the field theory. However, we shall see
that there is a one-to-one correspondence that allows us to compute the basic properties of the superfluid from microscopic physics, at least within our
simplifying assumptions of a uniform superflow and vanishing dissipation. The following 
points are crucial for this ``translation''.
\begin{itemize} 
\item The field-theoretic calculation is performed in the normal-fluid rest frame. In particular, ${\bf v}_s$, $\mu$, and $T$ are all measured in this frame.
The effective action density can be identified with the generalized pressure which in turn is related to the generalized energy density $\Lambda$. 

\item With the identification $\mu {\bf v}_s = -\nabla\psi$, $\mu = \partial^0\psi$, $T=\Theta^0$ and the condition for the normal-fluid rest frame $s^i=0$, 
eight independent variables are given in the microscopic approach. This number equals the number of degrees of freedom of two 4-vectors, for instance the 
two currents $j^\mu$ and $s^\mu$. We shall discuss how the other eight variables $s^0$, $\Theta^i$, $j^\mu$ as well as the coefficients $A$, $B$, $C$
can be computed as functions of ${\bf v}_s$, $\mu$, and $T$. 

\end{itemize}

In at least two important aspects, this translation between field theory and superfluid hydrodynamics is only a first step on which future work should build. 
Firstly, we only discuss stationary, homogeneous currents. From the hydrodynamic point of view this is only the simplest situation since the hydrodynamic 
equations $\partial_\mu j^\mu=\partial_\mu T^{\mu\nu}=0$ with the stress-energy tensor $T^{\mu\nu}$ are trivially fulfilled in this case. Secondly, since our microscopic
calculations are of thermodynamic nature, we do not include dissipation. Nevertheless, we may use our results for some nontrivial phenomenological 
features of superfluidity, most notably the two sound modes mentioned above. In particular, we can compute the velocities of first and second sound 
in the presence of a superflow (by superflow we always mean a {\it relative} flow between superfluid and normal fluid), see Fig.\ \ref{figsound}. 
\section{Condensate and superfluid velocity}
  
We start from the Lagrangian for a massless\footnote{See Ref.\ \cite{Alford:2012vn} for the case with a nonvanishing mass $m$. Since many expressions become quite 
lengthy when $m$ is included, we only discuss the case $m=0$ in these proceedings.}, complex scalar field $\varphi$ with quartic interactions,
\be \label{lagr}
{\cal L} = \partial_\mu\varphi\partial^\mu\varphi^* -\lambda|\varphi|^4 \, ,
\ee
with coupling constant $\lambda>0$. If the chemical potential associated to the conserved $U(1)$ charge is nonvanishing, there will be a Bose-Einstein condensate
which we parameterize by its modulus and its phase, $\langle\varphi\rangle = \frac{\rho}{\sqrt{2}}\, e^{i\psi}$. 
A chemical potential can be added by hand, but it can also be introduced through the phase of the condensate. In the latter case, 
the equations of motion allow for the following nonzero value of the condensate, 
\be \label{rho}
\rho = \frac{\sigma}{\sqrt{\lambda}} \, , \qquad \sigma\equiv \sqrt{\partial_\mu\psi\partial^\mu\psi} \, , 
\ee
which suggests that $\sigma$ plays the role of the chemical potential. In the simplest case, usually discussed in textbooks, bosons condense in the ground state
with vanishing four-momentum. Here we allow for a moving condensate, i.e., we want our bosons to condense in a state with 
non-vanishing momentum. This nonvanishing four-momentum is an external parameter and is identical to the four-gradient of the phase of the condensate $\partial^\mu\psi$.
In the following, $\rho$ and $\partial^\mu\psi$ are assumed to be constant in space-time. By equating the stress-energy tensor from its field-theoretic definition 
with its common hydrodynamic form, one finds that $\partial^\mu\psi$ is proportional to the superfluid four-velocity,
\be
v^\mu = \frac{\partial^\mu\psi}{\sigma} \, .
\ee 
Therefore, a constant $\partial^\mu\psi$ (and thus, via the equations of motion, a constant $\rho$) corresponds to a static, homogeneous superflow. It is important
to understand the meaning of the phase of the condensate and of $\sigma$: with the superfluid three-velocity 
${\bf v}_s  = -\nabla\psi/\partial_0\psi$ 
we can write $\sigma = \partial_0\psi\sqrt{1-{\bf v}_s^2}$. As indicated above, the Lorentz scalar $\sigma$ is the chemical potential, more precisely the chemical potential
in the frame where the superfluid velocity is zero, which we call the superfluid rest frame. 
We see that $\sigma$ differs from $\partial_0\psi$ only by a factor from a standard Lorentz transformation. Hence $\partial_0\psi$ is the chemical potential in the frame 
in which the condensate has velocity ${\bf v}_s$. Below we shall identify this frame with the normal-fluid rest frame in which we denote the chemical potential by $\mu$, i.e., 
we have $\mu = \partial_0\psi$. 
We thus see that topological modes of the phase play an important 
role: their winding numbers per unit time and per unit length around the $U(1)$ circle determine the chemical potential and the superfluid velocity, respectively. 
These topological modes are different from the topologically trivial oscillations around the ground state, which correspond to the collective excitations of the system.
By including small oscillations in the ansatz for the solution of the equations of motion, one obtains the dispersion relations for the two degrees
of freedom. In the presence of the condensate, these are the Goldstone mode and a massive mode. For small temperatures, only the Goldstone mode
becomes populated. We shall now discuss the thermodynamics of the system from the microscopic point of view.

\section{Goldstone mode and temperature expansion}

For sufficiently small temperatures (much smaller than the critical temperature) and with the assumption of uniformity, the effective action $\Gamma$ at the 
stationary point 
can be written as\footnote{For temperatures up to the critical temperature a more elaborate, self-consistent formalism is needed \cite{Alford:2007qa}.}
\be \label{effact}
\Gamma= -\frac{V}{T} \frac{\sigma^4}{4\lambda} - \frac{1}{2} \sum_k\Tr\ln \frac{S^{-1}(k)}{T^2} \, , \qquad S^{-1}(k) = \left(\begin{array}{cc} -k^2+2\sigma^2 &  2ik\cdot\partial\psi 
\\[2ex] -2ik\cdot\partial\psi & -k^2 \end{array}\right) \, , 
\ee
with the three-volume $V$, temperature $T$, the sum over four-momenta $k^\mu = (k_0,{\bf k})$, $k_0=-i\omega_n$ with the bosonic Matsubara frequencies 
$\omega_n$, and the inverse tree-level propagator in momentum space $S^{-1}(k)$. This expression is obtained by using the zero-temperature result for the condensate
(\ref{rho}). Effects of the melting of the condensate are 
neglected. The poles of the propagator, i.e., the zeros of the determinant of $S^{-1}(k)$, yield the dispersion relations. Since they are
solutions to a nontrivial quartic equation, they are very complicated in general. For small momenta, the dispersion of the Goldstone mode can be written as 
\be
\epsilon_{\bf k} =  c_1(x) |{\bf k}| + \frac{c_2(x)}{\mu^2} |{\bf k}|^3 + \ldots 
\ee
where we denote $x\equiv \cos\theta$ with $\theta$ being the angle between the superfluid velocity ${\bf v}_s$ and the momentum ${\bf k}$, and where
the coefficients in front of the linear and cubic terms are
\begin{subequations} \label{c1c2}
\bea
c_1(x) &=&  \frac{\sqrt{3-{\bf v}_s^2(1+2x^2)}\sqrt{1-{\bf v}_s^2}+2|{\bf v}_s|x}{3-{\bf v}_s^2} \, , \\[2ex]
c_2(x) &=&  \frac{(3-{\bf v}_s^2)^2+2{\bf v}_s^2x^2(3-10{\bf v}_s^2+3{\bf v}_s^4) -{\bf v}_s^4x^4(7-10{\bf v}_s^2-{\bf v}_s^4)}{(3-{\bf v}_s^2)^4\sqrt{1-{\bf v}_s^2}
\sqrt{3-{\bf v}_s^2(1+2x^2)}} \non[2ex]
&&-\,4|{\bf v}_s|x \frac{3-{\bf v}_s^2-{\bf v}_s^2x^2(1+{\bf v}_s^2)}{(3-{\bf v}_s^2)^4}  \, .
\eea
\end{subequations}
The massive mode has a mass $\epsilon_{{\bf k}=0} = \sqrt{2}\mu\sqrt{3-{\bf v}_s^2}$ and will not be relevant for the following since we assume $T\ll \mu$. As worked out in 
Ref.\ \cite{Alford:2012vn}, the coefficient of the linear term $c_1(x)$ is identical to the velocity of first sound, i.e., it can also 
be obtained from solving the sound wave equations, derived from the hydrodynamic conservation equations. The relevant wave equation equation can be written as
${\cal G}^{\mu\nu}k_\mu k_\nu=0$ with the so-called ``sonic metric'' ${\cal G}^{\mu\nu}\equiv g^{\mu\nu}+2v^\mu v^\nu$ \cite{Carter:1995if,Mannarelli:2008jq}. Moreover,
one can show that a Lorentz transformation of the four-vector $k^\mu = (c_1(x) |{\bf k}|, {\bf k})$ to the frame where the superfluid rests yields a transformed velocity
$c_1' = \frac{1}{\sqrt{3}}$, i.e., within our approximation the velocity of first sound is isotropic in the superfluid rest frame, even in the presence of a superflow.

For a translation into hydrodynamics and in particular into the two-fluid framework, 
we need to compute the stress-energy tensor and the current. With the help of the usual field-theoretic definitions we can write
\begin{subequations}
\bea
j^\mu &=& 
\partial^\mu\psi\,\frac{\sigma^2}{\lambda}- \frac{1}{2} \frac{T}{V}\sum_k\Tr\left[S\frac{\partial S^{-1}}{\partial (\partial_\mu\psi)}\right] \, , \label{jmu}
\\[2ex]
T^{\mu\nu} &=& \left(\partial^\mu\psi\partial^\nu\psi-g^{\mu\nu}\frac{\sigma^2}{4}\right)\frac{\sigma^2}{\lambda} 
- \frac{T}{V}\sum_k\Tr\left[S\frac{\partial S^{-1}}{\partial g_{\mu\nu}}-
\frac{g^{\mu\nu}}{2} \right] \,  . \label{dlnZ} 
\eea
\end{subequations}
For the stress-energy tensor we have used the gravitational definition, i.e., we take the derivative with respect to a general metric and return
to $g^{\mu\nu}=(1,-1,-1,-1)$ after taking this derivative. The evaluation of the effective action (\ref{effact}), the current and the stress-energy tensor is now
straightforward (up to a renormalization of $T^{\mu\nu}$, chosen such that the usual thermodynamic quantities in the absence of a superflow are obtained 
\cite{Alford:2012vn}). We refer the reader to Ref.\ \cite{Alford:2012vn} for all explicit results. As an example, we show the result for the effective action
density up to sixth order in temperature,
\be \label{tvgam}
\frac{T}{V}\Gamma \simeq \frac{\mu^4}{4\lambda}(1-{\bf v}_s^2)^2+\frac{\pi^2T^4}{10\sqrt{3}}\,\frac{(1-{\bf v}_s^2)^2}{(1-3{\bf v}_s^2)^2} 
- \frac{4\pi^4T^6}{105\sqrt{3}\,\mu^2}\,\frac{(1-{\bf v}_s^2)^2}{(1-3{\bf v}_s^2)^5}(5+30{\bf v}_s^2+9{\bf v}_s^4) \, .
\ee
For the $T^4$ term, knowledge of the slope of the linear part of the Goldstone mode
is sufficient, while the $T^6$ term knows about the linear {\it and} the cubic parts. Due to the superflow, the stress-energy tensor becomes anisotropic, i.e., 
there is a nonvanishing contribution $T^{0i}$ in the direction of ${\bf v}_s$, and the diagonal spatial elements have different longitudinal and transverse
projections with respect to ${\bf v}_s$, $T_\perp\equiv \frac{1}{2}\left(\delta_{ij}-\hat{\bf v}_{s,i}\hat{\bf v}_{s,j}\right)T_{ij}$, 
$T_{||}\equiv \hat{\bf v}_{s,i}\hat{\bf v}_{s,j} T_{ij}$. 
We find that the transverse projection is identical to the effective action density (\ref{tvgam}), $T_\perp = \frac{T}{V}\Gamma$. This is the generalization of the 
identity between the effective action density and the pressure in the isotropic case.
In terms of the two-fluid picture, the anisotropy of the stress-energy tensor is easily understood: in a single-fluid system one can always
choose a frame in which the stress-energy tensor is isotropic. For a two-fluid system, however, such a frame does not exist in general. In the rest frames of each of the 
fluid components, the pressure in the longitudinal direction with respect to the flow of the other fluid $T_{||}$ will differ from the pressure in the transverse direction
$T_\perp$. We now explain the two-fluid formalism for a nonzero-temperature superfluid.

\section{Covariant two-fluid formalism and sound velocities} 

In the superfluid hydrodynamics literature, it is often assumed that the starting point for the underlying microscopic physics is a ``master function''
\cite{Carter:1995if}
such as the so-called generalized pressure $\Psi$ which is a Lorentz scalar itself and a functional of the Lorentz scalars 
$\sigma^2$, $\Theta^2$, and $\Theta\cdot\partial\psi$. Each of these scalars is in general allowed to depend on the space-time coordinate $x$ (this dependence is 
absent in our simplified field-theoretic treatment above). A Legendre transformation from the conjugate momenta $\Theta^\mu$, $\partial^\mu\psi$ to the 
associated currents $j^\mu$, $s^\mu$ yields the generalized energy density $\Lambda  =-\Psi + j\cdot\partial\psi + s\cdot\Theta$, and the stress-energy tensor
is given by $T^{\mu\nu} = -g^{\mu\nu}\Psi+j^\mu\partial^\nu\psi+s^\mu\Theta^\nu$. With 
\be
\partial^\mu\psi = \frac{\partial\Lambda}{\partial j_\mu} \, , \qquad \Theta^\mu = \frac{\partial \Lambda}{\partial s_\mu}  \, , 
\ee
one obtains the momenta as linear combinations of the currents (\ref{psiTheta}) with the coefficients given by 
\be \label{ABC}
A\equiv \frac{\partial \Lambda}{\partial (j\cdot s)} \, , \qquad B\equiv 2\frac{\partial \Lambda}{\partial j^2}
\,, \qquad C \equiv 2\frac{\partial \Lambda}{\partial s^2} \, .
\ee
The connection with field theory is made by realizing that the generalized pressure is identical to the effective action density (and therefore identical to the 
transverse pressure $T_\perp$),
\be \label{PsiGamma}
\Psi = \frac{T}{V}\Gamma \, .
\ee
This can be read as a postulate. It also can be derived from the assumption that the field-theoretic calculation is performed in the normal-fluid rest 
frame \cite{Alford:2012vn}. This implies that the temperature of our statistical ensemble $T$ is measured in the normal-fluid rest frame. We identify 
$T=\Theta^0$. The relation (\ref{PsiGamma})
requires some explanation since we have just stated that $\Psi$ only depends on Lorentz scalars, while this is obviously not the case for the effective action which 
depends separately on $\partial^0\psi$, $\partial^i\psi$, and $\Theta^0$. Since we know how to compute the current $j^\mu$ [via (\ref{jmu})] and the entropy $s=s^0$ (via its thermodynamic 
definition), we may use Eqs.\ (\ref{psiTheta}) to express $A$, $B$, $C$ and the spatial components of $\Theta^\mu$ in terms of field-theoretic
quantities (in the normal-fluid rest frame, the spatial components of the entropy current vanish by definition, $s^i=0$). The results are 
\begin{subequations}
\bea
A=\frac{\partial^0\psi}{s^0{\bf j}\cdot\nabla\psi}\eta \, , \qquad B=-\frac{(\nabla\psi)^2}{{\bf j}\cdot\nabla\psi} \, ,  
\qquad     C= -\frac{j^0\partial^0\psi\,\eta -{\bf j}\cdot\nabla\psi s^0\Theta^0}{(s^0)^2\,{\bf j}\cdot\nabla\psi} \, , 
\eea
\end{subequations}
where we abbreviated $\eta\equiv {\bf v}_s^2j^0\partial^0\psi+{\bf j}\cdot\nabla\psi$, and 
\be
\Theta^i = -\frac{\partial^i\psi}{s^0}\left[j^0+\partial^0\psi\frac{{\bf j}\cdot\nabla\psi}{(\nabla\psi)^2}
\right] \, .
\ee
Again, one can evaluate these expressions in the small-temperature expansion, see Ref.\ \cite{Alford:2012vn} for the explicit results. One finds in particular that 
the entrainment coefficient $A$ vanishes for zero temperature which is expected since there is only a single fluid in this case. For any 
nonzero temperature, however, there is, in the language of the two-fluid model, an interaction between the two fluid components, characterised by 
$A$. One can also translate these coefficients into the normal-fluid and superfluid number densities $n_n$ and $n_s$ which are the more natural 
quantities if the system is decomposed in the spirit of the original, nonrelativistic two-fluid formalism, see Eq.\ (62) of Ref.\ 
\cite{Alford:2012vn}. 

With these relations between field theory and the covariant formalism we can also write down the generalized pressure in terms of Lorentz scalars for our 
particular microscopic model. To this end, we compute
\be
\sigma^2 = \mu^2(1-{\bf v}_s^2)\, , \qquad \Theta^2 \simeq \frac{(1-{\bf v}_s^2)(1-9{\bf v}_s^2)}{(1-3{\bf v}_s^2)^2}\,T^2   \, , \qquad 
\partial\psi\cdot\Theta \simeq \frac{1-{\bf v}_s^2}{1-3{\bf v}_s^2}\,\mu T  \, , 
\ee
where, in the last two relations, terms of order $T^3$ and $T^4$ are suppressed. This yields, up to order $T^4$, 
\be \label{Psonic}
\Psi \simeq \frac{\sigma^4}{4\lambda}+\frac{\pi^2}{90\sqrt{3}}
\left[\Theta^2+2\frac{(\partial\psi\cdot\Theta)^2}{\sigma^2}\right]^2 = \frac{\sigma^4}{4\lambda} + \frac{\pi^2}{90\sqrt{3}}({\cal G}^{\mu\nu}\Theta_\mu\Theta_\nu)^2\, , 
\ee
where we again recover the sonic metric ${\cal G}^{\mu\nu}$ which can be used for a compact notation of the $T^4$ term. In principle, our results also give the $T^6$ term. However, this term is, expressed in terms of Lorentz scalars, very complicated and we do not show its explicit form here. 

\begin{figure} [t]
\begin{center}
\hbox{\includegraphics[width=0.35\textwidth]{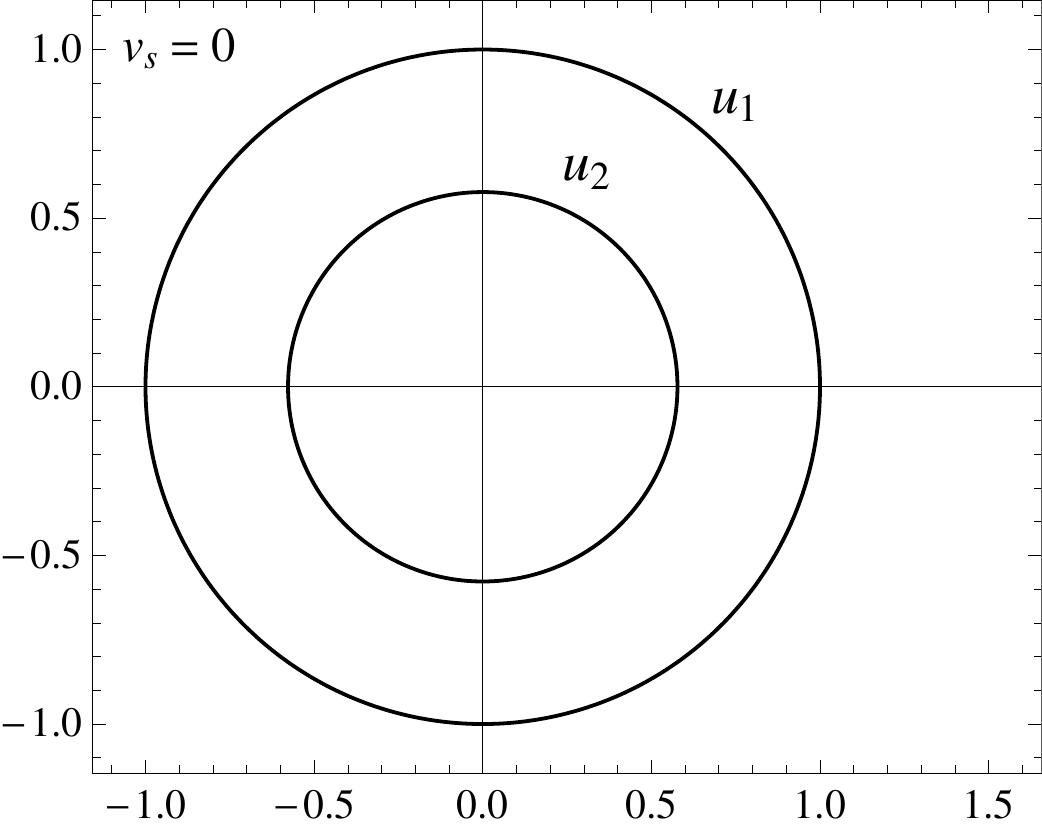}\includegraphics[width=0.319\textwidth]{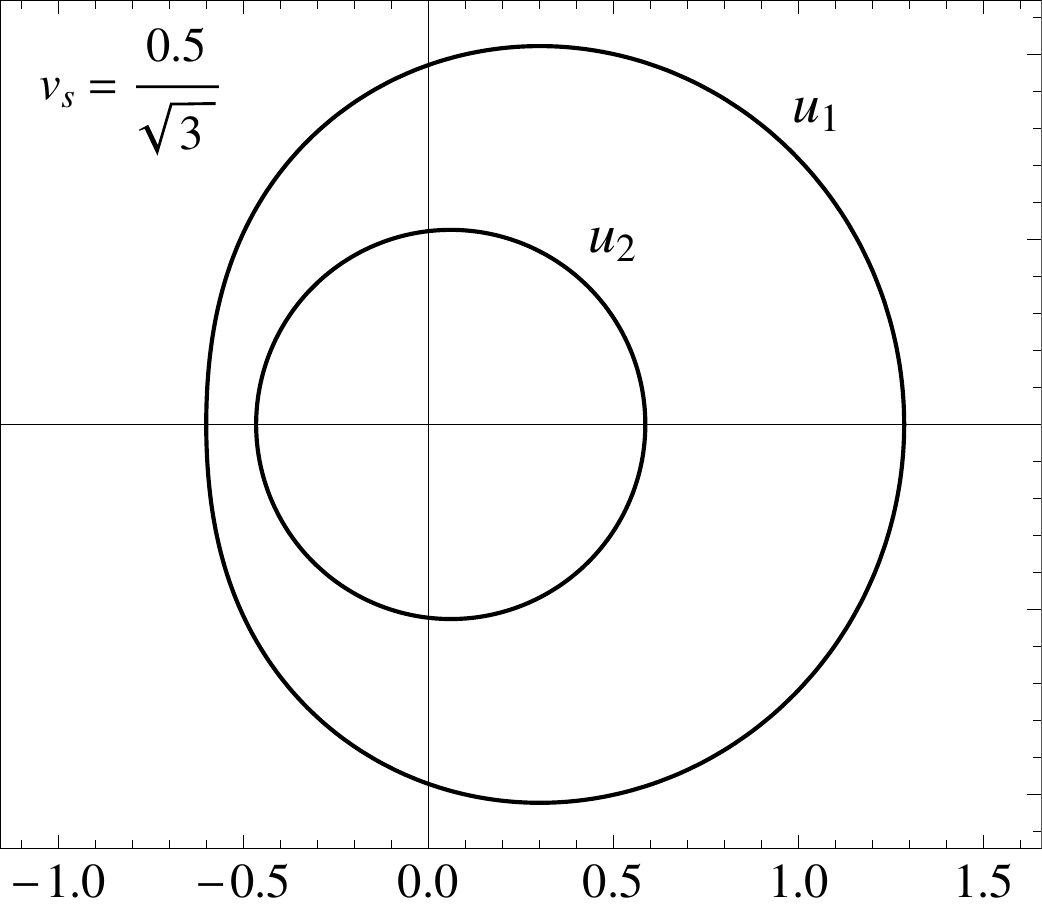}\includegraphics[width=0.319\textwidth]{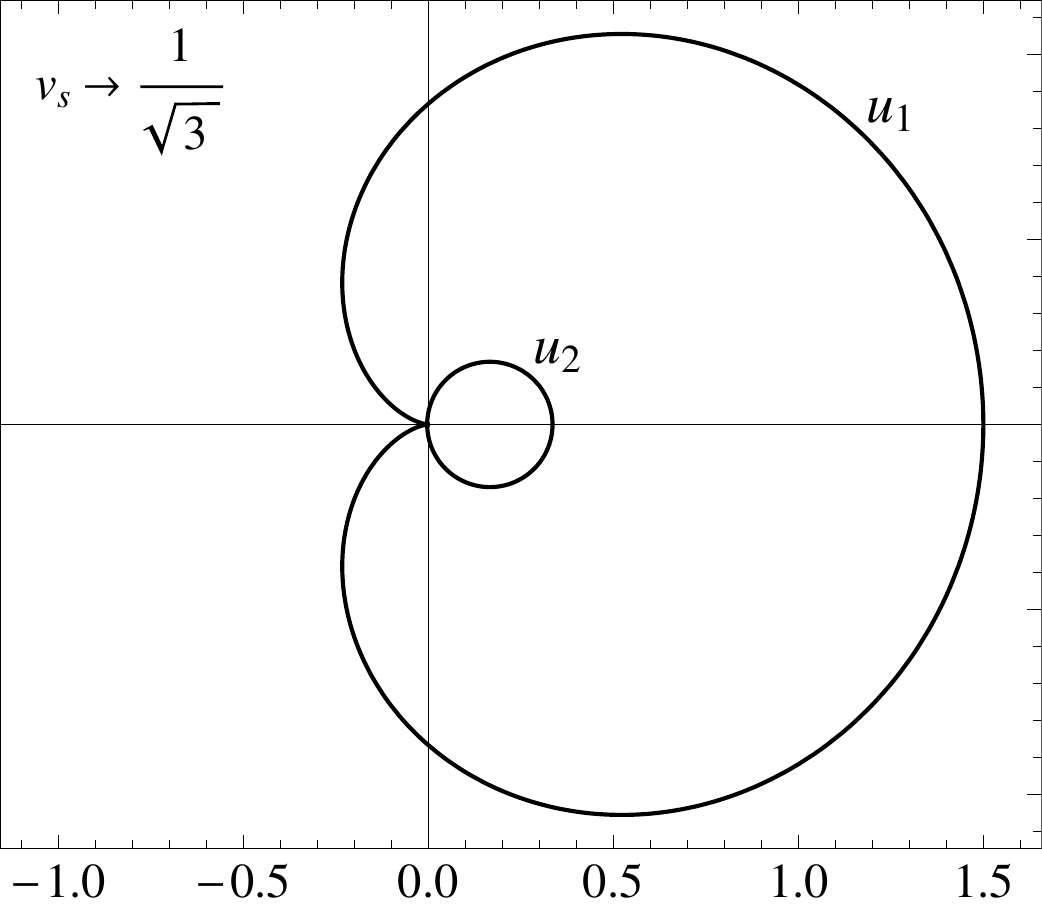}}
\caption{Polar plots of the velocities of first and second sound $u_1$, $u_2$ for zero temperature $T\to 0$ and for three different values of the superfluid 
velocity $|{\bf v}_s|$  (all velocities measured in the normal-fluid rest frame). 
The direction of the superflow is parallel to the horizontal axis and points to the right; the scale of the axes is normalized to the velocity of first sound in the absence 
of a superflow.} 
\label{figsound}
\end{center}
\end{figure}

We conclude with the presentation of the velocities of first and second sound, for which the above presented formalism can be used. The complete calculation 
can be found in Ref.\ \cite{Alford:2012vn}. The final results for $T\to 0$ are shown in Fig.\ \ref{figsound}. Both sound velocities
depend on the angle between the direction of the sound wave and the direction of the superflow. We find that the velocity of second sound decreases significantly
as the superflow approaches its critical value $|{\bf v}_s|=\frac{1}{\sqrt{3}}$ (beyond which the system becomes dissipative even at zero temperature). 
One can show that the velocity of second sound receives $T^2$ corrections for small temperatures which increase the velocity. In contrast, there is no 
temperature correction to the velocity of first sound within our small-temperature approximation \cite{Alford:2012vn}.

{\it Acknowledgments.} This work has been supported by the Austrian 
science foundation FWF under project no.~P23536-N16, and by U.S.~Department of Energy under contract
\#DE-FG02-05ER41375, 
and by the DoE Topical Collaboration 
``Neutrinos and Nucleosynthesis in Hot and Dense Matter'', 
contract \#DE-SC0004955.


 \end{document}